\documentstyle[graphicx,epsf,twocolumn,prl,aps]{revtex}

\begin{document}
\newlength{\figwidth}
\setlength{\figwidth}{3.375in}

\twocolumn[

\begin{center}
\large{\bf Calculation of quantum tunneling for a spatially extended
defect: the dislocation kink in copper has a low effective mass.}
\end{center}
\begin{center}
{
Tejs Vegge$^{*,\dag}$, 
James P. Sethna$^{\ddag}$,
Siew-Ann Cheong$^{\ddag}$, 
K.~W. Jacobsen$^{*}$,
Christopher R. Myers$^{\P}$
Daniel C. Ralph$^{\ddag}$,
\\[0.125\baselineskip]
{$^{*}$ Center for Atomic Scale Materials Physics (CAMP) and Department of
Physics,\\ Building 307, Technical University of Denmark, DK-2800, Kgs. Lyngby,
Denmark\\[0.125\baselineskip]
$^{\dag}$Materials Research Department, Ris{\o } National Laboratory
, DK-4000 Roskilde, Denmark\\[0.125\baselineskip]}
{$^{\ddag}$Laboratory of Atomic and Solid State Physics (LASSP), Clark Hall,
Cornell University, Ithaca, NY 14853-2501, USA\\[0.125\baselineskip]}
{$^{\P}$Cornell Theory Center,
Cornell University, Ithaca, NY 14853, USA\\[0.\baselineskip]}
}
\date{\today}
\end{center}

\begin{quote}
\begin{quote}\small
Several experiments indicate that there are atomic tunneling defects in
plastically deformed metals. How this is possible has not been clear, given
the large mass of the metal atoms.  Using a classical molecular-dynamics
calculation, we determine the structures, energy barriers, effective masses,
and quantum tunneling rates for dislocation kinks and jogs in copper screw
dislocations. We find that jogs are unlikely to tunnel, but the kinks should
have large quantum fluctuations. The kink motion involves hundreds of atoms
each shifting a tiny amount, leading to a small effective mass and
tunneling barrier.
\vspace*{0.5cm}
\end{quote}
\end{quote}
]


\vspace*{-1.5cm}

Tunneling of atoms is unusual.
At root, the reason atoms don't tunnel is that their
tunneling barriers and distances are set by the much lighter electrons. The
tunneling of a proton over a barrier one Rydberg high and one Bohr radius
wide is suppressed by the exponential of $\sqrt{2 M_p R_y a_0^2} = 
\sqrt{M_p / m_e} \sim 42.85$: a factor of $10^{-19}$. 

Nonetheless, atomic quantum tunneling dominates the low temperature
properties of glasses\cite{ZellerAHVP} and many doped
crystals\cite{Narayanamurti}. In glasses, there are rare regions (one
per $10^5$ or $10^6$ molecular units) where an atom or group of atoms
has a double well with low enough barrier, tunneling distance, and
asymmetry to be active. For certain dopants in crystals, off-center
atoms and rotational modes of nearly spherical ionic molecules have
unusually low barriers and tunneling distances. Quantitative modeling of
these spatially localized tunneling defects has been frustrated by the
demands for extremely accurate estimates of energy barriers, beyond even
the best density functional electronic structure calculations available
today. Also, although all detailed models of tunneling in glasses have
basically involved one or a very few atoms, there has long been
speculation that large numbers of atoms may be shifting during the
tunneling process.\cite{Ashcroft}

There is much evidence that quantum tunneling is important to the
properties of undoped, plastically deformed metals.  Quantum
creep\cite{Pustovalov}, glassy low-temperature behavior\cite{Thompson},
and two-channel Kondo scaling seen in the voltage and
temperature-dependent electrical conductivity in
nanoconstrictions\cite{RalphRefs} have been attributed to quantum
tunneling associated with dislocations. It has
never been clear how this can occur, given the large masses of the metal
atoms involved. 

We show here using a classical effective-medium
interatomic potential that quantum fluctuations can indeed be important
in the dynamics of one particular defect: a kink in the split-core screw
dislocation in copper. The motion of the kink involves a concerted
motion of hun-\break

\begin{figure}[thb]
\begin{center}
\vskip -0.3truein
\leavevmode
\epsfxsize=8cm
\epsffile{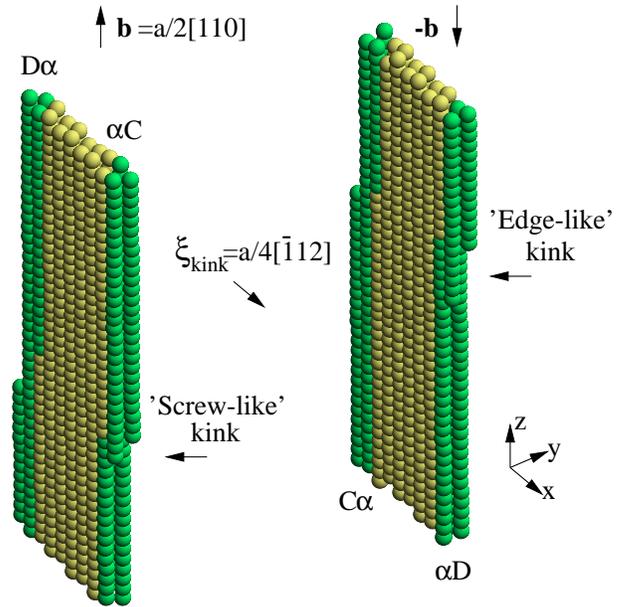}
\end{center}
\caption{{\bf Kink geometry.}
Broadly speaking, the screw dislocation represents the locus where
planes of atoms form a helix. In copper it spreads out into a ribbon along
the $x$-axis to lower its energy. The kink we study shifts the ribbon by
one atom in the x direction. 
More specifically, the {\bf b}=$\frac{a}{2}\left[110\right]$ dislocation on
the left dissociates
on the $(1\overline{1}1)$ plane into the Shockley partials:
$\alpha\mbox{C}=\frac{a}{6}[121]$
and
$\mbox{D}\alpha=\frac{a}{6}[21\overline{1}]$
respectively. The kink is introduced with line vector 
$\xi_{\mbox{{\small kink}}}=\frac{a}{4} [\overline{1}12]$, and 
dissociates into a wide screw-like- and a bulky edge-like kink
located on the partial dislocations. The lighter atoms are on the
stacking fault (hcp local environments) and the darker atoms are along
the partial dislocations (neither hcp nor fcc).
}
\label{fig:setup}
\end{figure}

\noindent 
dreds of copper atoms, leading to a dramatic decrease in
its effective mass.  This delocalization
perhaps lends support to ideas about collective centers in glasses. Also,
because our important conclusions rest upon this delocalization 
they are qualitatively much less sensitive to the accuracy of our potential than
calculations for spatially localized tunneling defects.
We assert that these kinks are likely the only
candidate for quantum tunneling in pure fcc metals. 


The kink simulation consists of two screw dislocations with opposite
Burgers vectors {\bf b}=$\pm\frac{a}{2}\left[110\right]$, allowing 
periodic boundary conditions giving us the perfect
translational invariance necessary to measure energy differences to 
the accuracy we need. The two dislocations are placed in different
$(1\overline{1}1)$ planes separated by 20 $(1\overline{1}1)$ planes (4.4 nm),
see figure \ref{fig:setup}. The system is 86 planes wide (19.3 nm) in the two
(non-orthogonal) directions, and extends 44.5 {\bf b} (11.4 nm) along the
dislocations. 
We introduce kinks or jogs on the dislocations by
applying skew periodic boundary conditions to system, {\em i.e.} we
introduce a small mismatch in the dislocation cores at the interface to
the next cell. The procedure also introduces a row of interstitial atoms
between the kinked dislocations, which is subsequently removed from the
system, leaving us with a total of 329\hspace{+0.05cm}102 atoms. The
kinks have a net line vector of $\xi_{\mbox{{\small
kink}}}=\frac{a}{4}[\overline{1}12]$, with $a$ the lattice constant.


To show how unusual the properties of the kink are, we also study the
properties of a dislocation jog.  The jog simulation, and the associated
energy barrier calculation, is similar and is described elsewhere\cite{VePe99}. 
The elementary jog we study is introduced with a line vector oriented in the
$(\overline{1}11)$ glide plane of the screw dislocation,
$\xi_{\mbox{{\small jog}}}=\frac{a}{4} [1\overline{1}2]$, which then
transforms into an obtuse lower energy configuration: 
$\frac{a}{4} [1\overline{1}2] \rightarrow \frac{a}{2} [101] +
\frac{a}{4} [\overline{11}0]$. This jog is
expected to be the most mobile of the jogs, second only to the kink in
mobility.

We introduce the two kinked dislocations directly as Shockley partial
dislocations, see
figure \ref{fig:setup}, and relax using the MD-min algorithm\cite{St97},
using Effective Medium Theory (EMT): a many-body classical
potential\cite{EMT}, which is computationally almost as fast
as a pair potential, while still describing the elastic properties well.
The elastic constants of the potential are: $C_{11}= 176.2 \ \mbox{GPa}$,
$C_{12}=116.0 \ \mbox{GPa}$ and $C_{44}=90.6 \ \mbox{GPa}$ with a Voigt
average shear modulus of $\mu=66\ \mbox{GPa}$, and an intrinsic stacking
fault energy of
$\gamma_{\mbox{\tiny I}}= 31 \ \mbox{mJ} / \mbox{m}^2$.

We present three quantities for the kink and jog: the Peierls-like barrier
for migration along the dislocation, the effective mass, and an upper bound
for the WKB factor suppressing quantum tunneling through that barrier.
Since the motion of these defects involves several atoms moving in a coordinated
fashion, we use instantons: the appropriate generalization of WKB analysis
to many-dimensional configuration
spaces\cite{Coleman,JPSThesis,PolyacetyleneTunneling}.
An upper bound for the
tunneling matrix element is given by the effective mass
approximation\cite{JPSThesis,SSH},
\begin{equation}
\Delta \le \hbar \omega_0
        \exp{\left(-\int \sqrt{2 M^*(Q) V^*(Q)} dQ / \hbar\right)},
\label{eq:UpperBound}
\end{equation}
where $\omega_0$ is an attempt frequency, $V^*(Q)$ is the energy of the defect
at position $Q$ with the neighbors in their relaxed, minimum energy positions
$q_i(Q)$, and 
\begin{equation}
M^*(Q) = \sum_i M_i (dq_i/dQ)^2
\label{eq:EffectiveMass}
\end{equation}
is the effective mass of the defect incorporating the kinetic energy of
the surrounding atoms as they respond adiabatically to its motion. The
effective mass approximation is usually excellent for atomic tunneling.
The method is variational, so equations \ref{eq:UpperBound} and
\ref{eq:EffectiveMass} remain upper bounds using other assumptions about
the tunneling path $q_i(Q)$ (such as the straight-line path between the
two minima described below for the kinks).

The difficulty of finding models for atomic tunneling is illustrated
rather well by the properties of the jog we study. The barrier for
migration was determined to be 15 meV\cite{VePe99}: lower than other
jogs, or even than surface diffusion barriers calculated with the same
potential. The effective mass for the jog, estimated by summing the
squared displacement of the 200 atoms with largest motion, is 
$M^{*}_{\mbox{\small jog}}\simeq 0.36 \ M_{\mbox{Cu}}$: the jog is spatially 
localized (it doesn't disassociate into partials), with a few atoms
in the core of the jog carrying most of the motion. The WKB tunneling matrix
element for the jog to tunnel a distance $Q=2.5$ {\AA } over a barrier 
$V=0.015 \ \mbox{eV}$ is suppressed by a factor of roughly
$\exp(-\sqrt{2 M^{*}_{\mbox{\small jog}} V} Q/\hbar) \simeq 10^{-14}$.
Jogs don't tunnel much.

\begin{figure}[thb]
\begin{center}
\leavevmode
\epsfxsize=8cm
\epsffile{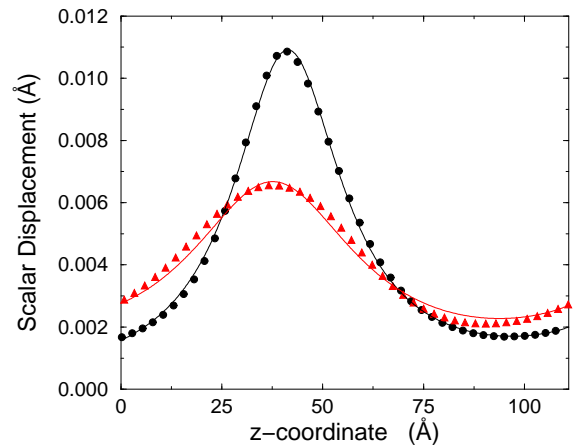}
\end{center}
\caption{The magnitude of the atomic displacement field as the two partial
kinks move, along the cores of the two partial dislocations, as a function
of the z-coordinate. The data points are fitted to 'periodic' Lorentzian
distributions. The FWHM value of the two partial kinks are 
$\Gamma_{\mbox{\small edge}}=13 {\bf b}$ (bullets)
and $\Gamma_{\mbox{\small screw}}=21 {\bf b}$ (triangles).
The center of masses of the two partial kinks are separated: the screw-like
partial kink is shifted by $\delta z=-1.5 {\bf b}$.}
\label{fig:profile}
\end{figure}

For the kinks, we take a relaxed initial configuration and define a
final configuration with the kink migrated by one lattice spacing along the
dislocation. 
The final position for each atom is given by the position 
of the neighboring atom closest to the current position minus the kink migration
vector $l_{\mbox{\small migr.}}=\frac{a}{2}\left[110\right]$ which represents
the net motion of the kink.
This automatically gives the correct relaxed final 
position, which is otherwise difficult to locate given the extremely small
barriers.
The width of the kinks is the traditional name for their extent along the
axis of the screw dislocation. We can measure this width
by looking at the net displacement of atoms between the initial and final
configurations. We find that the displacement field is localized into
two partial kinks, localized on the partial dislocation cores, see 
figure \ref{fig:profile}. These two partial kinks are quite wide (FWHM
of 13 {\bf b} and 21 {\bf b}). They differ because the partials are of
mixed edge and screw character; it is known\cite{HirthLothe} that the kink
which forces a mixed dislocation towards the screw direction will be
wider and have higher energy. This is wider than the $w < 10 {\bf b}$
predicted for slip dislocations in closed-packed
materials by Hirth and Lothe, and Seeger and Schiller using line tension
models \cite{Lothe-Hirth-Seeger-Schiller}.

\begin{figure}[thb]
\begin{center}
\leavevmode
\epsfxsize=8cm
\epsffile{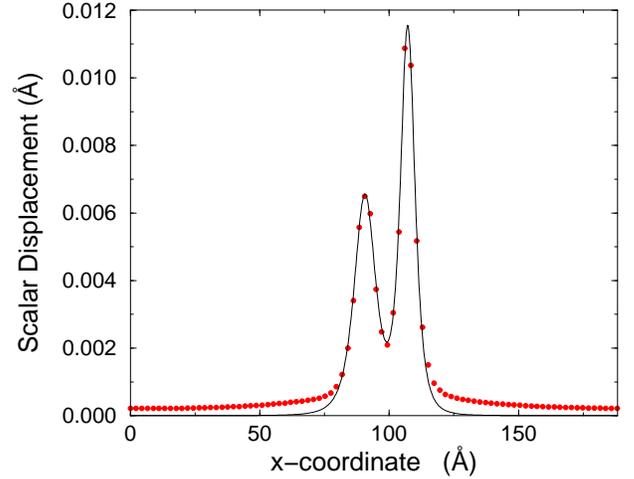}
\end{center}
\caption{The magnitude of the atomic displacement field as the two partial
kinks move, along the cores of the two partial dislocations, as a function
of the x-coordinate. The core regions are fitted to two squared Lorentzian
distributions. The partial core widths $W_{T} \simeq 4 d$ (4 $(1\overline{1}1)$ lattice planes), significantly reduces the effective mass of the kink.}
\label{fig:xWidth}
\end{figure}

Notice that the maximum net distance moved by an atom during the kink
motion in figure \ref{fig:profile} is around 0.01 \AA. Summing the
squares of all the atomic motions, and using equation
\ref{eq:EffectiveMass}, we find an effective mass $M_{\mbox{\small
kink}}^{*}\simeq M_{\mbox{\small Cu}} / 130$ within the straight-line
path approximation. This remarkably small mass can be attributed to
three factors. (1)~The mass is decreased because the screw dislocation is
 split into two partial dislocations \cite{Kroupa}.
(2)~The cores of the partial dislocations are spread transversally among
$W_T \simeq 4d$, figure \ref{fig:xWidth}; this factor seems to have been
missed in continuum treatments. These first two factors each reduce
the total distance moved by an atom as the kink passes from 
$z=-\infty$ to $+\infty$. (3)~The
kink partials average $W_L \simeq 17 {\bf b}$ wide (above), so the total atomic
motion is spread between around 17 kink migration hops\cite{EMWidth}.
Thus when the kink moves by $x$, the atoms in two regions $1/W_L$ long and
$1/W_T$ wide each move by $x/(2 W_L W_T)$, reducing\cite{EMWidth} the
effective mass by roughly a factor of $2 W_L W_T \sim 136$.

\begin{figure}[thb]
\begin{center}
\leavevmode
\epsfxsize=8cm
\epsffile{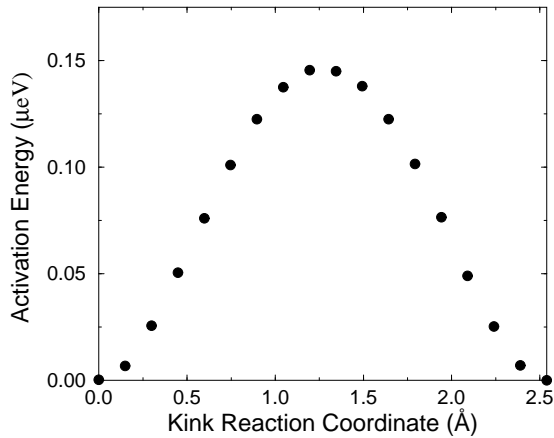}
\end{center}
\caption{The activation energy as a function of the straight-line distance
moved by the kink on one dislocation, with an associated barrier of
$E_{\mbox{\small act}} = 0.15 \ \mu \mbox{eV}$. Notice that this activation
energy is about one part in $10^{13}$ of the total system energy.
}
\label{fig:path}
\end{figure}

Evaluating the energy at equally spaced atomic configurations and linearly
interpolating between the initial and final states (along the
straight-line path) yields an upper bound to the kink-migration barrier
of $0.15\ \mu$eV, figure \ref{fig:path}. We attribute this extremely
small barrier to the wide kink partials: we expect the barrier $V$ to
scale exponentially with the ratio of the kink width $W$ to the (110)
interplanar distance {\bf b}: $V \sim \exp(-W/{\bf b})$. If one thinks of the
contribution to the energy of the $n^{\rm th}$ layer as some analytic
function $f(n/W + \delta)$, then the barrier is given by the variation
of the sum $\sum_{n=-\infty}^\infty f(n/W+\delta)$ with the position
shift $\delta$. The difference between this sum and the
($\delta$-independent) integral is easily estimated by Fourier transforms,
and is approximately $2 \tilde f(2 \pi W/a)$. The Fourier transform of an 
analytic function decays exponentially. One imagines this could be proven to
all orders in perturbation theory.

This small barrier is not only negligible for thermal activation (two mK),
but also for quantum tunneling. The WKB factor suppressing the tunneling
would be 
$\exp(-\sqrt{2 M^{*}_{\mbox{\small kink}} V} Q/\hbar) = \exp(-0.0148) = 0.985$.
Even at zero temperature, the kinks effectively act as free particles, as 
suggested in the literature (\cite{HirthLothe}, among others).

Our estimated kink migration barrier is thus $10^5$ times smaller than that
for the most mobile of the jogs. How much can we trust our calculation of
this remarkably small barrier?  Schottky\cite{Schottky} estimates using
a simple line-tension model that the barrier would be $\sim 3 \ 10^{-5}$ eV in 
fcc materials, using a Peierls stress $\sigma_{\mbox{\tiny P}} = 10^{-2}
\mu$ and a kink width $w=10$ {\bf b}. This value is a factor of 200 higher
than the barrier we find. On the other hand, both experiments and theoretical
estimates predict $\sigma_{\mbox{\tiny P}} \simeq 5 \ 10^{-6} \mu$ for
Cu \cite{Joos-Duesbery}, yielding barriers orders of magnitude lower than ours.
The interatomic potentials we use do not take into account directional bonding.
This is usually a good approximation for noble metals; however, small 
contributions from angular forces may change the kink width. The kink
width is like an energy barrier, balancing different competing energies 
against one another: in analogy, we expect it to be accurate to within
twenty or thirty percent. Our small value for the effective mass, 
dependent on the inverse cube of the spatial extent of the kink, 
is probably correct within a factor of two. The energy barrier is much more
sensitive: if we take the total exponential suppression to be $10^5$
(using the jog as a ``zero-length'' defect) then each 20\% change in the 
width would yield a factor of 10 change in the barrier height. The qualitative
result of our calculation, that the barriers and effective masses are
small, are robust not only to the use of an approximate classical potential,
but may also apply to other noble metals and perhaps simple and 
late transition metals.


In summary, we have used an atomistic calculation with classical potentials
to extract energy barriers and effective masses for the quantum tunneling
of dislocation jogs and kinks in copper. For jogs, the atomic displacements
during tunneling are primarily localized to a few atoms near the jog core,
each moving a significant fraction of a lattice spacing. Consequently,
the tunneling barrier and effective mass are relatively large, and 
tunneling is unlikely. However, the kinks in screw dislocations are 
much more extended: as a kink moves by one lattice spacing, hundreds of
atoms shift their positions by less than 1\% of a lattice spacing. Both
the energy barrier and the effective mass are reduced, to the extent that
tunneling should occur readily. Kinks are likely the only candidate
for quantum tunneling in pure crystalline materials. They may explain
measurements of quantum creep, glassy internal friction, and non-magnetic
Kondo effects seen in plastically deformed metals. 

\medskip \noindent {\bf Acknowledgments}. Portions of this work
were supported by the Digital Material project NSF \#9873214 and the
Danish Research Councils Grant \#9501775, and was done in collaboration with
the Engineering Science Center for Structural Characterization and Modeling
of Materials in Ris\o. An equipment grant from Intel and the support of the
Cornell Theory Center are also gratefully acknowledged. We had helpful
conversations with Nicholas Bailey and Jakob Schi{\o}tz.


\begin{references}

\bibitem{ZellerAHVP} R.C. Zeller and R.O. Pohl, {\sl Phys. Rev. B} {\bf 4},
2029 (1971).

\bibitem{Narayanamurti} V. Narayanamurti and R.O. Pohl, {\sl Rev. Mod. Phys.}
{\bf 42}, 201 (1970).

\bibitem{Ashcroft} K.~K. Mon and N.~W. Ashcroft, {\sl Solid State Comm.}
{\bf 27}, 609 (1978).

\bibitem{Pustovalov} B.V.~Petukhov and V.L.~Pokrovskii, JETP Lett.
\textbf{15}, 44 (1972) calculate the quantum rate for double kinks
using a simple line tension model; V.V.~Pustovalov, Fiz.~Nizk.~Temp. {\bf 15},
901 (1989), translated in Sov. J. Low Temp.\ Phys., gives a review;
A. Hikata and C. Elbaum, {\sl Phys. Rev. Lett.} {\bf 54}, 2418 (1985).

\bibitem{Thompson} T.~Kosugi, D.~McKay and A.V.~Granato, J.~Phys.~IV
\textbf{6}, 863 (1996), X. Liu, E-J. Thompson, B.E. White, Jr.,
and R. O. Pohl, {\sl Phys. Rev. B} {\bf 59}, 11767 (1999), and to be published
in the 1999 Boston MRS symposium L.

\bibitem{RalphRefs} D.C. Ralph, A.W.W. Ludwig, J. von Delft, and R.A. Buhrman,
{\sl Phys. Rev. Lett.} {\bf 72}, 1064 (1994); R.J.P. Keijsers,
O.I. Shklyarevskii, and H. van Kempen, {\sl Phys. Rev. Lett.} {\bf 77}, 3411
(1996); G. Z{\'a}rand {\it et al.}, {\sl Phys. Rev. Lett.} {\bf 80}, 1353
(1998); O.P. Balkashin {\it et al.}, {\sl Phys. Rev. B} {\bf 58}, 1294
(1998); S.K. Upadhyay {\it et al.}, {\sl Phys. Rev. B} {\bf 56}, 12033 (1997).

\bibitem{VePe99} T.~Vegge, O.~B.~Pedersen, T.~Leffers and K.~W.~Jacobsen
(to appear in Mat. Res. Soc. Symp. Proc.)

\bibitem{St97} P.~Stoltze, {\em Simulation Methods in Atomic-scale Materials
    Physics} (Polyteknisk forlag, 1997).

\bibitem{EMT} K.W.~Jacobsen, J.K.~N{\o}rskov, and M.J.~Puska, {\sl Phys.
 Rev. B}, {\bf 35}, 7423, (1987).  K.W.~Jacobsen, P.~Stoltze, and 
J.K.~N{\o}rskov, {\sl Surf. Sci.} {\bf 366}, 394, (1996), and references
therein.

\bibitem{Coleman} S. Coleman, 1977 Erice lectures, reprinted in his 
{\sl Aspects of Symmetry}, Cambridge Univ. Press, Cambridge, 1985, p. 265.

\bibitem{JPSThesis} James P. Sethna, {\sl Phys. Rev. B} {\bf 24}, 698 (1981),
{\bf 25}, 5050 (1982).

\bibitem{PolyacetyleneTunneling} James P. Sethna and Stephen Kivelson, 
{\sl Phys. Rev. B Rapid Communications} {\bf 26}, 3513 (1982).

\bibitem{SSH} W.P. Su, J.R. Schrieffer, and A.J. Heeger, {\sl Phys. Rev. Lett.}
{\bf 42}, 1698 (1979); {\sl Phys. Rev. B} {\bf 22}, 2099 (1980).

\bibitem{HirthLothe} Hirth and Lothe, {\em Theory of Dislocations} second ed.,
Krieger, Malabar Fl., 1992: quantum kink tunneling p. 533ff.

\bibitem{Lothe-Hirth-Seeger-Schiller} J.~Lothe, J.~P.~Hirth,
Phys.~Rev. \textbf{115}, 543 (1959), and A.~Seeger
and P.~Schiller, in \emph{Physical Acoustics}, edited by W.~P.~Mason,
Academic Press (New York), 1966, Vol.~3A, pp.~361; J.~D.~Eshelby, 
Proc.~R.~Soc.~London Ser.~A \textbf{266}, 222 (1962). 

\bibitem{Kroupa} F.~Kroupa and L.~Lej\v{c}ek, Czech.~J.~Phys.~B
\textbf{25}, 65 (1975)

\bibitem{EMWidth} Hirth and Lothe \protect\cite{HirthLothe} 
constant line tension p. 252.
A.~Seeger and P.~Schiller, and 
J.~D.~Eshelby\protect\cite{Lothe-Hirth-Seeger-Schiller} 
recognize
that the effective mass is reduced by a factor of the longitudinal width of
the kink, which Hovakimian {\it et al.} use to calculate depinning of
dislocation kinks (L.B.~Hovakimian, Phys.~Rev.~Lett. \textbf{68},
1152 (1992), L.B.~Hovakimian, K.~Kojima and I.~Okada, Phys.~Rev.~B
\textbf{51}, 6319 (1995)).

\bibitem{Schottky} G.~Schottky, Phys.~Stat.~Sol.~\textbf{5}, 697 (1964)
  
\bibitem{Joos-Duesbery} B.~Jo\'os, M.~S.~Duesbery,
  Phys.~Rev.~Lett. \textbf{78}, 266 (1997), and references therein.

%
%
%
%
%
%
%

\end{references}
\end{document}